**Comment on 'Unveiling the double-well energy landscape in a ferroelectric layer'**

J. A. Kittl[1,2], M. Houssa[2], V. V. Afanas'ev[2], J. -P. Locquet[2]

[1]Advanced Logic Lab, Samsung Semiconductor Inc., Austin TX 78754, USA

[2]Department of Physics and Astronomy, K. U. Leuven, Leuven, Belgium

For over 10 years, many papers have been published that assume it is possible to stabilize negative capacitance in a ferroelectric layer by capacitance matching to an adjacent dielectric layer [1-8]. These publications claim that under such stabilization, the ferroelectric polarization vs electric field trajectory is a stable reversible trajectory with an "s-shape" typically referred as the "s-curve" [1-8], rather than the common hysteretic trajectory observed generally in ferroelectrics [9]. Recently, the paper "Unveiling the double-well energy landscape in a ferroelectric layer" by M. Hoffmann, et al., Nature 565, 464 (2019) [8], presented an experiment which was claimed to demonstrate directly for the first time such stabilization. Indeed, the paper presents an "s-curve" extracted from measurements performed on a ferroelectric-dielectric bilayer capacitor, and also claimed non-hysteretic cycling. We show here, simply by plotting the data obtained from [8] and based on all information provided in that reference and to our knowledge, that both claims lack foundation and are incorrect, i.e. that the trajectory of the ferroelectric is hysteretic and does not follow an s-curve, showing that, based on the data and information presented in ref. [8] the s-curve presented in [8] is simply an artificial extraction from the data.

Recently, the underlying models [1-6] postulating negative capacitance stabilization in a ferroelectric by capacitance matching to a dielectric have been criticized and postulated to be unphysical (i.e. in contradiction with basic physical principles) [9], and alternative explanations have been provided for experimental data [9-11], such as transient effects (transient negative capacitance) and ferroelectric switching hysteresis compensated by charge trapping-detrapping hysteresis [9-14]. However, recently, the paper "Unveiling the double-well energy landscape in a ferroelectric layer" by M. Hoffmann, et al., Nature 565, 464 (2019) [8], presented an experiment which was claimed to demonstrate directly for the first time a measurement revealing the "s-curve" for a ferroelectric-dielectric (FE-DE) bilayer capacitor as well as claims of lack of hysteresis in cycling. We first point out that small signal capacitance measurements performed for a similar FE-DE bilayer capacitor presented by the same group of authors in [7] showed no evidence of stabilized negative capacitance (actually ruling it out): if the polarization-electric field trajectory of the FE was indeed a stable reversible 's-curve", then this trajectory should be followed also in the small signal experiments. The authors then performed pulsed measurements with pulse durations in the μs time scales. Initially, a large negative voltage was applied to the FE-DE bilayer capacitor (top electrode) to ensure the polarization of the ferroelectric was already switched to negative polarization (pointing up). A train of positive voltage pulses is then applied (pulse durations in the μs time scales) with peak voltages increasing gradually each pulse, and returning to 0 voltage in between pulses. Fig. 1a shows the charge-voltage plots of the FE-DE bilayer capacitor, obtained from the data from [8]. The charge is the integral of the current - throughout the whole pulsed measurement experiment-, and is the total capacitor charge corrected for the initial charge at the beginning of the pulsed measurements, $Q_{cap_0}$ (charge after the application of the large negative voltage and voltage reset to 0, before starting the sequence of pulses). At low peak voltages, only the dielectric response is seen. The plots in Fig. 1a use a series resistance of 350 ohms (as used in ref [8]). As peak voltage increases, some ferroelectric domains start to switch polarization. Since the peak voltage is increased gradually, only a few domains switch per minor loop, and the charge increment per minor loops is small. However, it accumulates through the experiment showing a normal hysteretic behavior (given by the envolvent in Fig. 1a). We see that the claim of lack of hysteresis in [8] is not correct. The data shows normal hysteretic behavior. A significant issue in the analysis presented in [8] is that the charge is calculated as the integral of the current over each pulse separately (i.e. not considering the charge accumulated in previous pulses): the charge is artificially reset to 0 at the beginning of each pulse.

Next, we analyze the "construction" of the s-curve presented in Fig. 3h of [8]. We first recall that the s-curve is supposed to be –according to stabilized NC models- a stable reversible trajectory of the ferroelectric polarization vs electric field. We first note that, as explained in [9], the negative capacitance portion of the s-curve cannot be derived from the Landau theory of ferroelectrics to be part of a stable trajectory as proposed in [8]. Although the whole s-curve can be derived as the locus of points with zero first derivative of the Landau free energy of a ferroelectric (under some conditions) as described in [8], it is easy to verify that the region of negative capacitance corresponds to maxima of free energy (still with 0 first derivative), which cannot represent physically stable states, while the rest of the curve corresponds to minima (physically stable states).

In any case, the authors propose a method to calculate the polarization and electric field within the ferroelectric. We noted several issues with the calculation of these quantities. We already mentioned the fact that residual charge accumulated in previous pulses (and related to FE switching) was ignored. Also, the authors ignored the fact that any change in charge in the plates of the capacitor results in a change in voltage across the dielectric layer (unless charges were injected during each pulse from the plates into the stack, which would render many of the calculations invalid in any case), and assumed only the reversible charges affect the voltage across the dielectric layer (note that with FE polarization switching, residual charges accumulate which impact the voltage across the dielectric as well). Furthermore, biases were introduced artificially in both axis shifting the curve that was calculated to center it on 0 polarization and 0 field (ensuring it looks like an s-curve), thus the fact that the constructed curve passes through the origin is forced, not a conclusion of the experiment. However, the major criticism to ref. [8] is that it did not show the minor loops of the experiment, rather, it presented a locus of points constructed using only one point per minor loop. Although non-optimal, we used the exact methodology used in the data analysis in [8], but instead of just calculating one point per minor loop we plot in Fig. 1b the whole minor loops for a few pulses (for clarity). These are the polarization-electric field trajectories of the FE -as calculated by the methods proposed in [8]-. They do not resemble s-curves. They could be consistent with minor loops exhibiting transient negative capacitance observed in FE capacitors (although with some arbitrary shifts on both axis). We also highlighted in black full circles the points taken to fabricate the s-curve, corresponding roughly to the maximum values of the polarization for each pulse (corresponds to maximum value of what the authors in [8] call "reversible charge"). The artificial connection of these points gives the claimed s-curve, seen here to not correspond to any trajectory of the FE.

This analysis shows that 1) the claim of lack of hysteresis in ref [8] is incorrect, since the experiments show clear FE hysteresis, and 2) that the s-curve constructed in [8] has no physical meaning as a stable trajectory, should not be connected to a Landau model (which in any case assigns no meaning to the negative capacitance portion of proposed s-curves), and, if anything, is simply a collection of isolated points from normal minor loops in a ferroelectric capacitor switching under conditions that result in transient negative capacitance, which are artificially connected to create the impression of a trajectory.

**Figures**

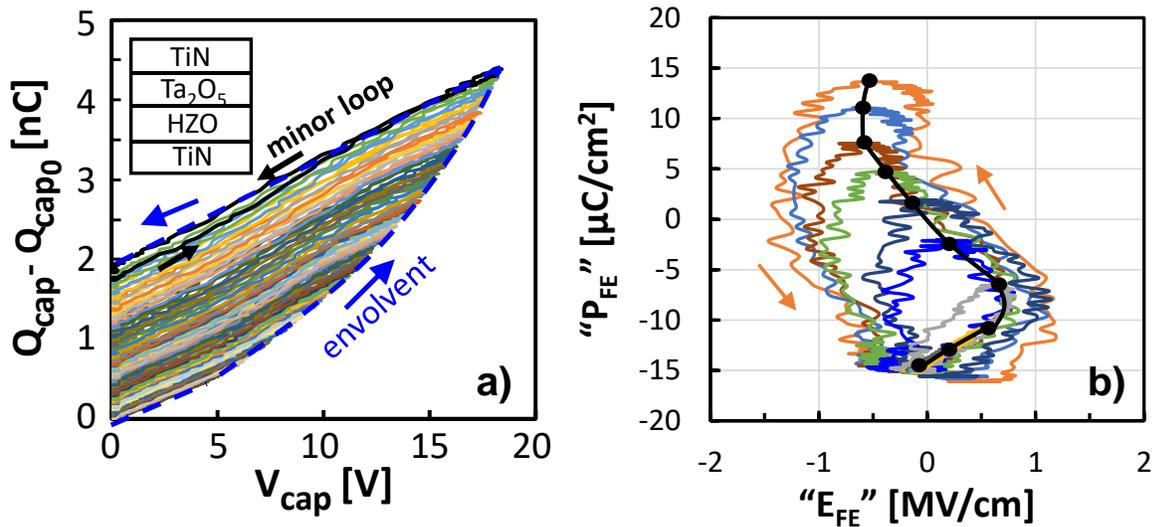

**Figure 1.** a) Charge-voltage trajectory for a DE-FE bilayer capacitor from data in ref. [8], showing conventional FE hysteresis. The experiment starts with a large negative voltage ensuring all domains are switched in the same direction, and resulting in an initial capacitor charge $Q_{cap_0}$ (at 0 V, at the start of the pulse train). A sequence of µs range pulses of increasing voltage is applied. As peak pulse voltage is increased, some domains in the FE start to switch polarization resulting in charge accumulation. Since the peak voltage increment (from the peak voltage of the previous pulse) for each pulse is small, only few domains switch in each pulse, but the effect is cumulative. Normal hysteretic behavior is observed for each pulse (minor loops) and the overall FE hysteresis (charge accumulation through the experiment) is depicted by the envolvent (blue dotted line). b) Trajectory of the quantities labelled as "$P_{FE}$" and "$E_{FE}$" in ref. [8] calculated by the expressions given in ref. [8] (which includes centering arbitrarily the curves in both axis) with data from ref. [8], for several pulses in the experiment (different colors represent different poulses). For each pulse only one point of the trajectory was plotted in ref. [8] (indicated as black dots in the plot), the connection of these points for different pulses (shown as a black line) was presented as the "s-curve" trajectory of the ferroelectric, despite the fact that the actual trajectories do not resemble the s-curve, and that the calculations forced this curve to go through the origin of the plot (artificially centered).